# Drift-induced Unidirectional Graphene Plasmons

Tiago A. Morgado[1], Mário G. Silveirinha[1,2*]

[1]*Instituto de Telecomunicações and Department of Electrical Engineering, University of Coimbra, 3030-290 Coimbra, Portugal*

[2]*University of Lisbon, Instituto Superior Técnico, Avenida Rovisco Pais, 1, 1049-001 Lisboa, Portugal*

E-mail: tiago.morgado@co.it.pt, mario.silveirinha@co.it.pt

**Abstract**

Nonreciprocal photonic devices enable "one-way" light flows and are essential building blocks of optical systems. Here, we investigate an alternative paradigm to break reciprocity and achieve unidirectional subwavelength light propagation fully compatible with modern all-photonic highly-integrated systems. In agreement with a few recent studies, our theoretical model predicts that a graphene sheet biased with a drift electric current has a strong nonreciprocal tunable response. Strikingly, we find that the propagation of the surface plasmon polaritons can be effectively "one-way" and may be largely immune to the backscattering from defects and obstacles. Furthermore, the drift-current biasing may boost the propagation length of the graphene plasmons by more than 100%. Our findings open new inroads in nonreciprocal photonics and offer a new opportunity to control the flow of light with one-atom thick nonreciprocal devices.

---

[*] To whom correspondence should be addressed: E-mail: mario.silveirinha@co.it.pt

Light propagation in conventional photonic systems is constrained by the Lorentz reciprocity law. This fundamental principle is intimately related to the invariance of Maxwell's equations under time-reversal symmetry (*1*), which forbids one-way light flows in standard metal-dielectric platforms. According to the reciprocity principle (*2-3*), if the positions of the source and receiver are interchanged the level of the received signal remains the same. Thus, reciprocal systems are inherently bidirectional.

With the ever-increasing demand for all-photonic highly-integrated systems (*4*), there has been recently a tremendous effort in the development of solutions that permit nonreciprocal light propagation. The standard way to break the Lorentz reciprocity principle is by using a static magnetic field bias that creates a gyrotropic nonreciprocal response (*5-10*). In particular, it was recently demonstrated that some gyrotropic material platforms are intrinsically topological, and thereby may support unidirectional scattering-immune edge states (*11-20*). However, the need of an external magnetic biasing hinders the integration of such components in nanophotonic systems. Furthermore, the gyrotropic response is usually rather weak at optics because for a realistic magnetic bias strength the cyclotron frequency typically lies in the microwave range. Transistor-loaded metamaterials provide a viable path to create a strong nonreciprocal response with no static magnetic fields, but only at microwave and millimeter-wave frequencies (*21-22*).

Alternative solutions to obtain asymmetric light flows have been extensively investigated, namely by exploiting nonlinear effects (*23-27*) and opto-mechanical interactions (*28-31*). Nevertheless, the requirement of high power signals in nonlinear systems and the typically weak response of opto-mechanical resonators limit the applicability of these solutions. Spatio-temporally modulated waveguides also offer interesting conceptual opportunities to break the reciprocity of light propagation, but these approaches may be difficult to implement and integrate in nanophotonic systems (*32-36*).

In this Letter, we explore a simple and innovative platform that provides magnetic-free nonreciprocal subwavelength light propagation through the biasing of a graphene sheet with a drift electric current (Fig. 1). Even though the electric drift current bias is a well-established solution to break the Lorentz reciprocity (*37-38*), it received only marginal attention in the recent literature because significant nonreciprocal effects require large drift velocities, which are impracticable in metals and in most semiconductors. Here, we highlight that graphene may offer a truly unique opportunity in this context: it has an ultra-high electron mobility and enables drift velocities on the order of $c/300$ (*39-43*), which may be several orders of magnitude larger than in typical metals (*38*) and several times larger than in high-mobility semiconductors (*44-45*). Motivated by these properties, we theoretically show that the high electron mobility of graphene can enable a broadband subwavelength "one-way" propagation regime and waveguiding immune to backscattering due to the presence of near-field scatterers or other structural imperfections. Remarkably, the drift-current biasing can change the graphene conductivity dispersion and boost the propagation length of the graphene plasmons.

A few recent works studied the impact of an electric current bias in the graphene response in different contexts (*46-50*), but the possibility of waveguiding immune to back-reflections in graphene was not discussed. The drift current bias was taken into account in (*46-48*) through a semiclassical correction of the equilibrium distribution of the electrons in graphene, but its effect on the energy dispersion and wave functions of the electronic states was neglected. In contrast, the conductivity model adopted in the present study is obtained using purely quantum mechanical methods (self-consistent field approach; see the supplementary materials of Ref. (*50*)). The effect of the drift current is described by an approximate interaction Hamiltonian of the form $\hat{H}_{\text{int,drift}} = \mathbf{v}_0 \cdot \hat{\mathbf{p}}$ ($\hat{\mathbf{p}} = -i\hbar\nabla$ is the momentum operator), which shifts the electrons velocity by $\mathbf{v}_0$. This interaction Hamiltonian neglects the dependence of the drift velocity on the electron energy. Different from the theory of (*46-48*), our approach predicts the frequency Doppler shift due to the motion of the drifting electrons and hence stronger nonreciprocal effects. We would like to underscore that a drift current bias is fundamentally different from the standard electrical tuning of the graphene response through the control of the Fermi level, see e.g., (*51-52*).

- **Results and Discussion**

To begin with, we investigate the propagation of the SPPs supported by a graphene sheet traversed by a DC electric current supplied by a voltage generator. In the absence of a drift current, the graphene sheet is characterized by a surface conductivity $\sigma_g(\omega)$, which is described by the Kubo formula and takes into account both intraband and interband transitions (*53-55*). We assume throughout this manuscript that the chemical potential of the graphene sheet is $\mu_c = 0.1$ eV, $T = 300$ K (i.e., room temperature), and that the intraband and interband scattering rates are $\Gamma_{\text{intra}} = 1/(0.35 \text{ ps})$ and $\gamma_{\text{inter}} = 1/(0.0658 \text{ ps})$, respectively (*54*). The graphene sheet is surrounded by a dielectric (SiO$_2$ or h-BN). Moreover, it is assumed that the time variation is of the form $e^{-i\omega t}$, where $\omega$ is the oscillation frequency. For simplicity, here we neglect the intrinsic nonlocal response of graphene (*55-56*) ($\sigma_g = \sigma_g(\omega, k_x)$), but its impact on the SPP propagation is analyzed in the Supporting Information for low temperatures.

Recently, it was demonstrated by us (*50*) using the self-consistent field approach that in the presence of drifting electrons with velocity $\mathbf{v} = v_0 \hat{\mathbf{x}}$ the graphene conductivity associated with a longitudinal excitation (with in-plane electric field directed along *x*) is of the form $\sigma_g^{\text{drift}}(\omega, k_x) = (\omega/\tilde{\omega})\sigma_g(\tilde{\omega})$, where $\tilde{\omega} = \omega - k_x v_0$ is the Doppler-shifted frequency and $k_x$ is the wave number along the *x*-direction. Interestingly, this result implies that with the drift current (i.e., for $v_0 \neq 0$) $\sigma_g^{\text{drift}}(\omega, k_x) \neq \sigma_g^{\text{drift}}(\omega, -k_x)$, and thereby the electromagnetic response of the graphene sheet is nonreciprocal (*57*). Note that the time-reversal operation flips the drift velocity, and thus the response of the graphene sheet with a drift current is not time-reversal invariant (*1*). It is worth mentioning that temperature gradients can also create drift currents, and thereby offer a different mechanism to break the reciprocity of graphene.

To study the opportunities created by the drift current biasing, next we characterize the graphene plasmons. By solving Maxwell's equations, it can be shown that the dispersion characteristic of the SPPs supported by a graphene sheet is determined by $\frac{2}{\gamma_s} i\omega\varepsilon_0\varepsilon_{r,s} - \sigma_g^{\text{drift}} = 0$ (*55*), where *c* is the speed of light in vacuum, $\varepsilon_0\varepsilon_{r,s}$ is the permittivity of the dielectric substrate, and $\gamma_s = \sqrt{k_x^2 - \varepsilon_{r,s}(\omega/c)^2}$ is the transverse (along *z*) attenuation constant that determines the confinement of the graphene plasmons. Evidently, without the drift current the same formula holds with $\sigma_g^{\text{drift}}$ replaced by $\sigma_g$.

Figure 2A depicts the dispersion characteristic of the SPPs for different drift velocities $v_0$. For clarity, we represent both the positive and negative frequency solutions. The reality of the electromagnetic field implies that the positive and negative frequency solutions are mapped one into another as $\omega(k_x) \to -\omega(-k_x^*)$ (particle hole-symmetry). Because of the graphene material absorption the guided wave number of the SPPs is typically complex-valued, $k_x = k_x' + ik_x''$. In the absence of the drift-current ($v_0 = 0$) the $\omega$ vs. $k_x$ dispersion curve is formed by two symmetric branches [black solid and dashed curves in Fig. 2A] corresponding to two counter-propagating waves. The two positive frequency branches are linked by $\omega(k'_x) = \omega(-k'_x)$, in agreement with the reciprocity and parity symmetries of the system. In contrast, with the drift current biasing, there is an evident symmetry breaking of the SPPs dispersion such that the $+x$ and $-x$ directions become inequivalent. The asymmetry of the dispersion curve is more pronounced when the drift velocity increases. A similar but weaker effect was predicted in Refs. (*46-48*). A detailed comparison between our findings and those of previous works is reported in the Supporting Information. Notably, the symmetry breaking gives rise to a wide frequency range wherein the SPPs are allowed to propagate along the $+x$ direction (the direction of the drift velocity), but are forbidden (or strongly attenuated) in the opposite direction. For instance, for a drift velocity $v_0 = v_F / 2$ ($v_F \approx c/300$ is the Fermi velocity), a regime of unidirectional propagation along the $+x$ direction emerges for frequencies above 10 THz [see the blue solid curves of Fig. 2A]. Even though the drift velocity in graphene can be on the order of the Fermi velocity (*39-43*), it is interesting to see from Fig. 2A that the one-way propagation regime may be attainable for drift velocities as low as $v_0 = v_F / 8$. Furthermore, as illustrated in Fig. 2B the graphene plasmons that propagate along the negative *x*-direction are more strongly affected by dissipation than plasmons that propagate along the positive *x*-direction. The inclusion of nonlocal corrections in the graphene response weakens somewhat the nonreciprocal response, but does not change the unidirectional character of the plasmons (see the Supporting Information), namely the discrepancy between the attenuation constants $|k''_x|$ of counter-propagating graphene plasmons. Nonlocal effects are stronger when $|k'_x|$ is comparable to or larger than the Fermi wave number $k_\text{F} = \mu_c / (\hbar v_\text{F})$ ($k_\text{F} = 0.15$ nm$^{-1}$ in Fig. 2) (see the Supporting Information), but such a region is not interesting for waveguiding applications due to the relatively short propagation distance of short-wavelength SPPs.

To further characterize the properties of the graphene plasmons with a drift-current biasing, we calculated the propagation length $\delta_x = 1/\text{Im}\{k_x\}$ as a function of the lateral decay length $\delta_z = 1/\text{Re}\{\gamma_s\}$. Figure 2C shows the ratio between the SPP propagation lengths ($\delta_x^\text{drift} / \delta_x^\text{no-drift}$) with and without a drift-current biasing for the $k_x' > 0$ branch represented with solid lines in Fig. 2A. Remarkably, the propagation length of the SPPs with the drift current may significantly exceed that of the SPPs in the absence of a bias current. For example, for $v_0 = v_\text{F}/2$ and $\delta_z = 10$ nm the distance travelled by the SPP wave is twice as large as for the case without biasing. Indeed, for the lateral confinement $\delta_z = 10$ nm, the SPP propagation length is $\delta_x = 174$ nm for $v_0 = v_\text{F}/2$, whereas in the absence of a bias current the propagation length is only $\delta_x = 85$ nm. The increased propagation length of the SPPs stems from the Doppler shift undergone by the graphene conductivity due to the drifting electrons, which changes the material dispersion and weakens the dissipation effects. It should be noted that due to the drift-current bias our system has an "active" response (*50*), which may play some role in the enhanced propagation length. Unlike the bilayer graphene system considered in (*50*), the configuration under study does not lead to electromagnetic instabilities and spasing because such an effect requires the interaction of two surface modes.

To further highlight the emergence of a one-way propagation regime in the drift-current biased graphene, next we consider a scenario wherein the plasmons are excited by a linearly polarized emitter placed in the close proximity of the graphene sheet (Fig. 3A). The radiated and scattered electromagnetic fields are calculated using the formalism described in the Supporting Information. Figures 3B-D show time snapshots of the longitudinal component of the electric field ($E_x$) excited by the near field of the emitter. In the first example (Fig. 3B), it is assumed that there is no drift current. Clearly, in this case the wave propagation is bidirectional and two identical counter-propagating SPPs are excited by the emitter [see Fig. 3B]. Quite differently, when a biasing DC current is used, the response of the system becomes strongly nonreciprocal and the SPPs are "dragged" by the drifting electrons so that they can propagate only along the $+x$ direction [see Fig. 3C-D], in agreement with Fig. 2A. Qualitatively similar results are obtained when the nonlocality of the bare graphene response is taken into account (see the Supporting Information).

Here, we note that other works have previously demonstrated the excitation of directive plasmons in graphene (*58-61*). However, such proposals rely on asymmetric excitations of the SPPs, e.g., by using gratings (*58-60*) or emitters with circular polarization (*61*). These solutions are fully reciprocal and hence the graphene plasmons are bidirectional. In contrast, our system is genuinely "one-way", and thereby independently of the excitation or of the polarization of the emitter, the plasmons are launched towards the $+x$-direction.

The "one-way" property implies that the graphene plasmons cannot be backscattered by an obstacle or defect, somewhat similar to topological structures (*6, 11-15*). In our system the dielectric is a transparent material. Hence, in principle the plasmons may be scattered by an obstacle as plane-wave states of the dielectric, even though for near-field interactions such a process is inefficient. Moreover, since graphene is a lossy material, part of the energy transported by the plasmon may be dissipated near the obstacle. To illustrate these ideas, next we study how the presence of scatterers affects the SPP propagation. We consider a setup similar to that of Fig. 3A, except that one or more metallic strips are placed along the SPPs propagation path [see Fig. 4A]. For simplicity all the obstacles are characterized by the same relative permittivity ($\varepsilon_{r,i} = -4.1$) and by the same width ($w_i = 12$ nm). To determine the influence of the objects on the wave propagation we use the formalism described in the Supporting Information, which treats the obstacles as secondary sources of radiation.

The calculated time snapshots of the *x*-component of the electric field are depicted in Fig. 4B-C. Without a drift current, the propagation is bidirectional, and for this reason the obstacles strongly backscatter the SPP mode [see Fig. 4B*i-ii*]. Quite differently, with a biasing drift current (Fig. 4C*i-ii*) there is no backward channel of propagation (i.e., along $-x$ direction), and hence no back-reflection is observed. In this case, the plasmons are forced to go around the obstacles and a strong transmission level is observed. As previously noted, part of the energy of the plasmons may be dissipated at the obstacles or leak into the bulk dielectric. From the density plots, the latter process appears to be of secondary importance.

The backscattering suppression is even more evident from the results of Fig. 4B*iii*-C*iii*, where $E_x$ is depicted as a function of *x* and for $z = 0$. As seen, when the graphene sheet is under a DC current biasing (Fig. 4C*iii*), the field before the obstacles is unperturbed, even when the propagation path is obstructed by two scatterers [purple dot-dashed curve in Fig. 4C*iii*]. On the other hand, beyond obstacles position, the field amplitude is almost the same as when the propagation path is free. The slight amplitude reduction is mainly due to material absorption. In contrast, without the bias drift current (Fig. 4B*iii*), the forward SPP is back-reflected so strongly by the obstacles that it practically vanishes beyond the obstacles positions. A more detailed analysis of the influence of the obstacles on the transmitted field is given in the Supporting Information.

- **Conclusions**

In summary, we have shown that a drift current breaks the reciprocal response of the graphene conductivity and enables a broadband regime of "one-way" SPP propagation. The proposed one-atom thick nonreciprocal platform has a tunable and switchable response: the graphene conductivity can be controlled either by changing the drift velocity or the chemical potential. Furthermore, the direction of propagation of the unidirectional plasmons is determined by the sign of the drift velocity. Importantly, the drift-current biasing enhances the propagation length of the graphene SPPs by factors of up to 2-4. The unidirectional SPPs are protected against backscattering from obstacles and defects, analogous to topological systems. The proposed system can be used as a building block of a wide range of nonreciprocal circuits, e.g., optical isolators and circulators. Thus, our findings open up a new and exciting route to control the flow of light in highly-integrated nonreciprocal tunable nanophotonic circuits.

**Supporting Information**

The Supporting Information reports (*A*) the comparison of our theory with other results from the literature and the analysis of the impact of nonlocal effects; (*B*) derivation of the plane-wave reflection and transmission coefficients for a drift-current biased graphene sheet; *(C)* and *(D)* derivation of the fields radiated by a linearly polarized emitter placed near the drift-current biased graphene without and with scattering obstacles, respectively; (*E*) study of the attenuation caused by the interaction of an obstacle with the SPP wave. This material is available free of charge via the Internet at http://pubs.acs.org.

**Acknowledgments:** This work was partially funded by Fundação para a Ciência e a Tecnologia PTDC/EEI-TEL/4543/2014 and by Instituto de Telecomunicações under project UID/EEA/50008/2017. T. A. Morgado acknowledges financial support by Fundação para a Ciência e a Tecnologia (FCT/POPH) and the cofinancing of Fundo Social Europeu under the Post-Doctoral fellowship SFRH/BPD/84467/2012.

# Figures

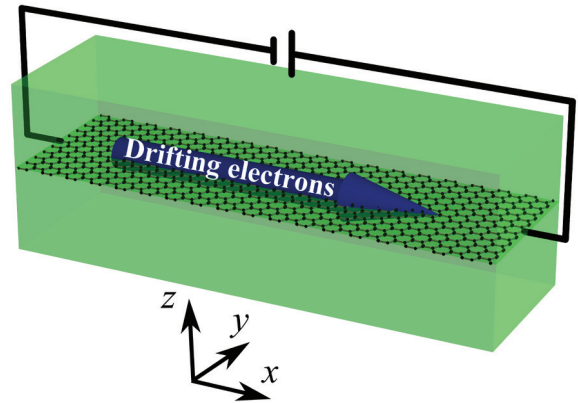

**Fig. 1. Drift current biasing of graphene.** A static voltage generator induces an electron drift in a graphene sheet embedded in a dielectric with relative permittivity $\varepsilon_{r,s} = 4$.

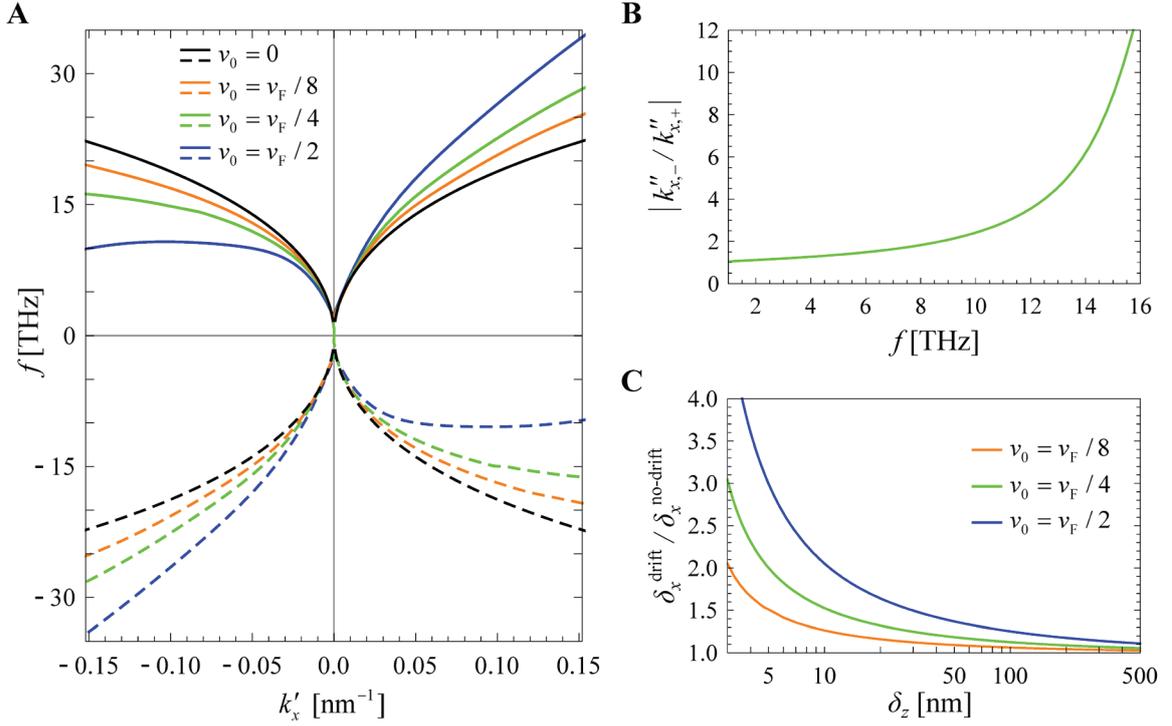

**Fig. 2. SPPs dispersion diagram.** (A) Dispersion of the SPPs supported by the graphene sheet for different drift velocities $v_0$ and for both positive and negative frequencies. (B) Ratio between the imaginary parts of the guided wave numbers of the drift-induced SPPs that propagate along the $-x$ and $+x$ directions as a function of the frequency for a drift velocity $v_0 = v_F/4$. (C) Ratio between the SPP propagation lengths ($\delta_x$) with and without drift-current biasing as a function of the SPP lateral decay length ($\delta_z$) for the forward SPP (propagating along $+x$; the SPP dispersion is depicted with a solid line in panel A).

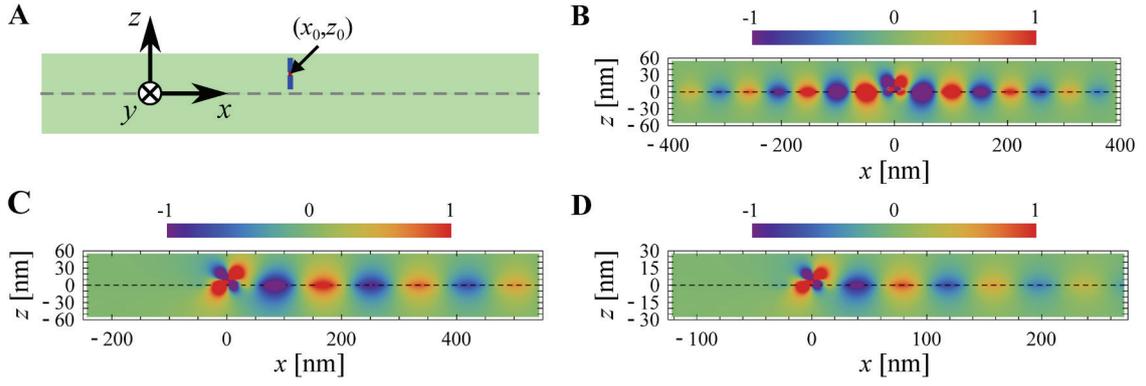

**Fig. 3. SPP excitation by a near-field emitter.** (**A**) A linearly polarized emitter (with vertical polarization) is placed near the graphene sheet at $(x,z) = (0, z_0)$. (**B-D**) Time snapshots of the $x$-component of the electric field $E_x$ (in arbitrary unities). (**B**) Without a drift current bias. (**C-D**) With a drift current bias. (**B**) $v_0 = 0$, $z_0 = 10$ nm, and $f = 15$ THz; (**C**) $v_0 = v_F/2$, $z_0 = 10$ nm, and $f = 15$ THz; (**D**) $v_0 = v_F/4$, $z_0 = 5$ nm, and $f = 20$ THz.

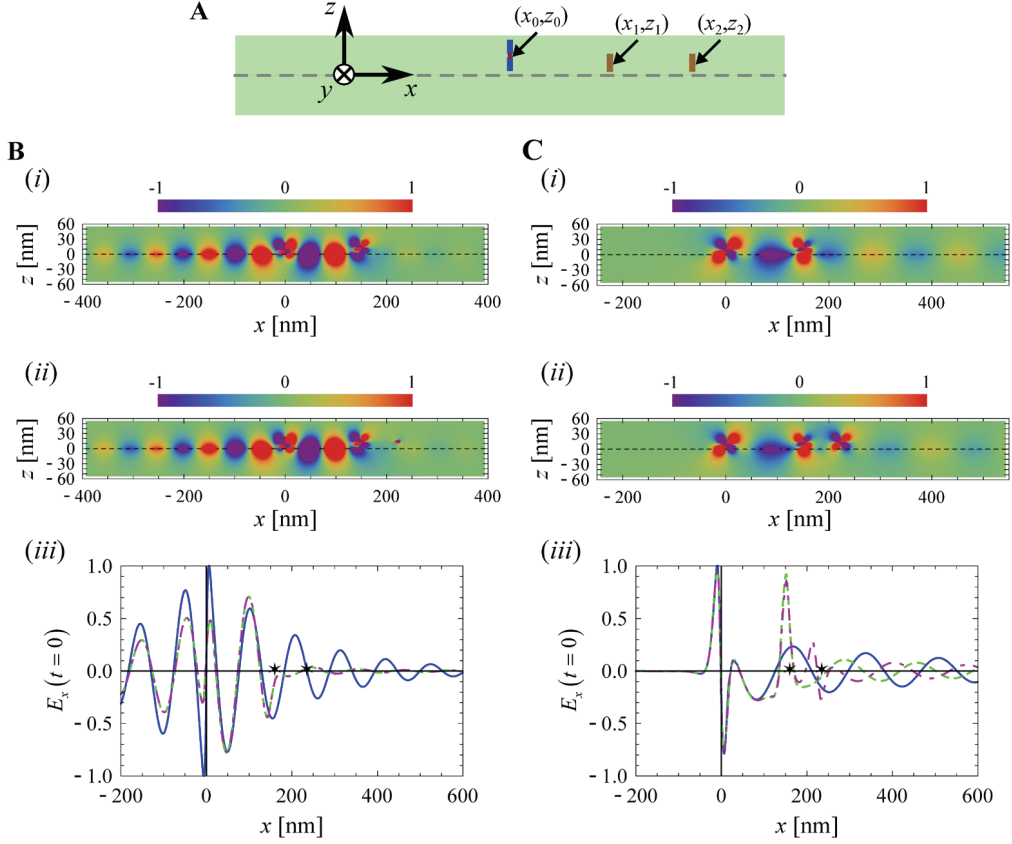

**Fig. 4. SPP propagation with obstacles.** (**A**) A linearly polarized emitter and thin strip obstacles are placed near the graphene sheet. (**B-C**) Similar to Fig. 3B-C, but with one or more obstacles at the positions $(x_i, z_i)$. (**B**) Without a drift current bias. (**C**) With a drift current bias with $v_0 = v_F/2$. (**B***i*-**C***i*) With a single obstacle at $(x_1, z_1) = (150\text{ nm}, 15\text{ nm})$; (**B***ii*-**C***ii*) With two obstacles at $(x_1, z_1) = (150\text{ nm}, 15\text{ nm})$ and $(x_2, z_2) = (225\text{ nm}, 15\text{ nm})$. (**B***iii*-**C***iii*) Snapshots in time of $E_x$ (in arbitrary unities) as a function of $x$ and for $z = 0$. Blue solid lines: without obstacles; green dashed lines: with a single obstacle; purple dot-dashed lines: with the two obstacles. The two black stars in (**B***iii*-**C***iii*) represent the position of the obstacles; the emitter is at the origin.

# Supporting Information for

# "Drift-induced Unidirectional Graphene Plasmons"

Tiago A. Morgado[1], Mário G. Silveirinha[1,2*]

[1]*Instituto de Telecomunicações and Department of Electrical Engineering, University of Coimbra, 3030-290 Coimbra, Portugal*

[2]*University of Lisbon, Instituto Superior Técnico, Avenida Rovisco Pais, 1, 1049-001 Lisboa, Portugal*

*E-mail:* tiago.morgado@co.it.pt, mario.silveirinha@co.it.pt

In the supplementary note *A*) we study the impact of nonlocal effects and compare our theory with results of other works. In the supplementary note *B*), we find the plane-wave reflection and transmission coefficients for a drift-current biased graphene sheet. In the supplementary notes *C)* and *D)* we derive the electromagnetic fields radiated by a linearly-polarized emitter placed above a graphene sheet without scattering objects, and with scattering objects, respectively. Finally, in supplementary note *E)* we study how the obstacle attenuation varies with the frequency and with the drift velocity.

## A. *Impact of nonlocal effects and comparison with other works*

Following our previous work (*1*), the nonlocal effects in the bare graphene conductivity ( $\sigma_g = \sigma_g(\omega, k_x)$ ) can be taken into account simply by considering that the conductivity with a drift current is:

$$\sigma_g^{\text{drift}}(\omega, k_x) = \frac{\omega}{\tilde{\omega}} \sigma_g(\tilde{\omega}, k_x), \tag{S1}$$

where $\tilde{\omega} = \omega - k_x v_0$ is the Doppler-shifted frequency. The function $\sigma_g(\omega, k_x)$ (i.e., the nonlocal bare graphene conductivity) is calculated as explained in Refs. (*2-5*), but the relevant integrals are difficult to evaluate. Fortunately, a closed form expression for

---

$\sigma_g(\omega, k_x)$ is available in the zero-temperature limit. Specifically, the $T=0$ nonlocal conductivity is given by Eqs. (S1a) and (S1d) of the supporting material of Ref. (*6*). The approximation $\sigma_g \approx \sigma_g(\omega, k_x = 0) \equiv \sigma_g(\omega)$ yields the standard (local) Kubo formula. The formalism of the main text (based on the approximation $\sigma_g^{drift}(\omega, k_x) \approx (\omega/\tilde{\omega})\sigma_g(\tilde{\omega})$) neglects the nonlocal effects in the bare graphene response.

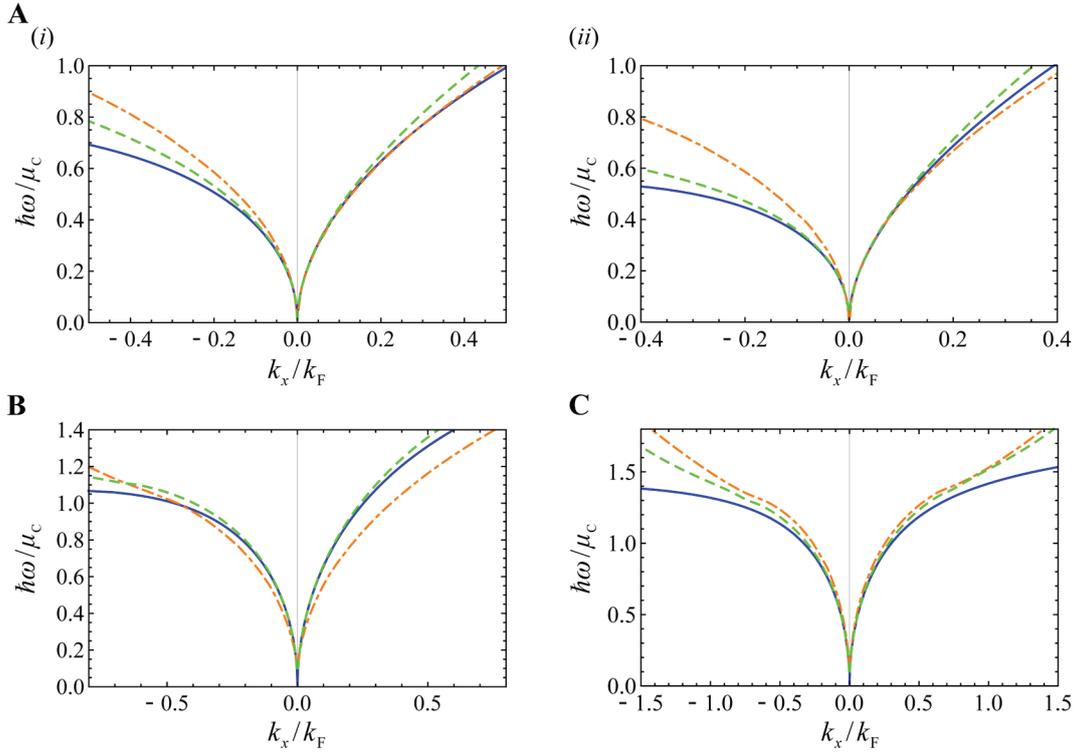

**Fig. S1.** Dispersion of the SPPs supported by the drift-current biased graphene sheet calculated using three different methods. Blue solid lines: our model with a local $\sigma_g$; Green dashed lines: our model with a nonlocal $\sigma_g$; Orange dot-dashed lines: results reported in **(A)** Ref. (*7*), **(B)** Ref. (*8*), and **(C)** Ref. (*9*). **(A)** $\varepsilon_{r,s} = 2.5$ and (*i*) $v_0 = 0.3 v_F$, (*ii*) $v_0 = 0.6 v_F$; **(B)** $v_0 = 0.3 v_F$ and $\varepsilon_{r,s} = 1.0$; **(C)** $v_0 = 0.05 v_F$ and $\varepsilon_{r,s} = 1.0$. The calculations are done for $T = 0$ K and the effect of collisions is neglected. $k_F = \mu_c/(\hbar v_F)$ is the Fermi wavenumber.

Figure S1 shows the dispersions of the drift-induced graphene SPPs calculated with our formalism including (green dashed lines) or neglecting (solid blue lines) the nonlocal effects in the bare graphene response superimposed with the results reported in Refs. (*7-9*) [orange dot-dashed lines]. On the overall, one can see that even though the inclusion of the nonlocal effects weakens somewhat the nonreciprocal response, it does not change the unidirectional nature of the SPPs. Indeed, broadband regimes of unidirectional SPP propagation can emerge even when the nonlocal effects are taken into account (see Fig. S1A*ii*). For large drift velocities, the semiclassical models from (*7-9*) [orange dot-dashed lines] predict considerably weaker nonreciprocal responses than our "local" and "nonlocal" results. In fact, the degree of asymmetry between the SPPs associated with $+k_x$ and $-k_x$ is clearly smaller for the results of Refs. (*7-9*). Hopefully, future experiments will help understanding which of the theories models better the drift current bias.

Figure S2 depicts a time snapshot of the *x*-component of the electric field radiated by a linearly-polarized emitter above a graphene sheet biased with drifting electrons with $v_0 = v_F / 2$, including the correction due to the nonlocal bare graphene response.

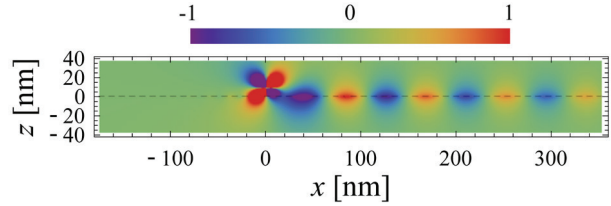

**Fig. S2.** Time snapshot of the electric field $E_x$ (in arbitrary unities) when a linearly polarized emitter (with vertical polarization) is placed near the graphene sheet at $(x,z)=(0,10\,\text{nm})$. The graphene is biased with drifting electrons with $v_0 = v_F/2$ and the nonlocal corrections of the bare graphene conductivity are taken into account in the calculation. The remaining parameters are $f = 25\,\text{THz}$, $T = 0\,\text{K}$, and $\Gamma_{\text{intra}} = 1/(0.17\,\text{ps})$ (*8*).

As seen, the results are qualitatively analogous to those of Fig. 3C-D of the main text, and the emitted wave propagates only towards the +*x*-direction. The emitted field is computed using the same formalism as in the main text but taking into consideration the nonlocal corrections of the bare graphene response [Eq. S1].

## B. Reflection and transmission coefficients

The reflection and transmission coefficients for a TM-polarized wave incident on a drift-current biased graphene sheet (see Fig. S3) can be obtained in the usual way by expanding the electromagnetic field in all the regions of space in terms of plane waves, and then solving for the unknown wave amplitudes with mode matching.

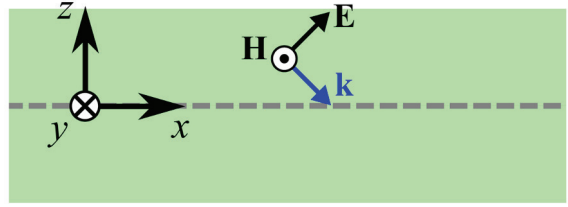

**Fig. S3.** Sketch of a graphene sheet (with a drift-current bias) surrounded by dielectric with relative permittivity $\varepsilon_{r,s}$.

Assuming that the incident magnetic field is along the $y$-direction and has complex amplitude $H_y^{inc}$, it follows that the magnetic field in all space can be written as (the variation $e^{i(k_x x - \omega t)}$ of the fields is omitted):

$$H_y = H_y^{inc}\left(e^{\gamma_s z} + \rho e^{-\gamma_s z}\right), \quad z > 0$$
$$H_y = H_y^{inc} \tau e^{\gamma_s z}, \quad z < 0 \tag{S2}$$

The electric field associated with a generic plane wave $\mathbf{H} = H_0 e^{i\mathbf{k}\cdot\mathbf{r}}\hat{\mathbf{y}}$ is $\mathbf{E} = \dfrac{1}{-i\omega\varepsilon_0\varepsilon_{r,s}}\nabla\times\mathbf{H}$, and hence it is straightforward to decompose the electric field into plane waves, similar to Eq. (S2). By matching the tangential component of the electric field ($E_x\big|_{z=0^+} - E_x\big|_{z=0^-} = 0$) and by imposing the impedance boundary condition ($H_y\big|_{z=0^+} - H_y\big|_{z=0^-} = -\sigma_g^{drift} E_x$) at the interface (*2, 10*), it is found that the reflection and transmission coefficients satisfy

$$\rho(k_x,\omega) = \frac{\gamma_s}{\gamma_s - 2\varepsilon_{r,s}\kappa_g^{drift}}, \quad \text{and} \quad \tau(k_x,\omega) = -\frac{2\varepsilon_{r,s}\kappa_g^{drift}}{\gamma_s - 2\varepsilon_{r,s}\kappa_g^{drift}}, \tag{S3}$$

where $\sigma_g^{\text{drift}}(\omega, k_x) = (\omega/\tilde{\omega})\sigma_g(\tilde{\omega})$ is the graphene conductivity in the presence of a drift-current bias, $\tilde{\omega} = \omega - k_x v_0$ is the Doppler-shifted frequency, $\sigma_g(\omega)$ is the standard graphene conductivity given by the Kubo formula (2, 10) and $\kappa_g^{\text{drift}} = i\omega\varepsilon_0/\sigma_g^{\text{drift}}$. Evidently, when $v_0 = 0$ the reflection and transmission coefficients reduce to the standard formulas in the absence of a drift-current bias (2).

## C. Fields radiated by a linearly polarized emitter above a graphene sheet

We suppose that the emitter (polarized along the vertical direction) is characterized by the current density $\mathbf{j}_e = -i\omega p_e \delta(x-x_0)\delta(z-z_0)\hat{\mathbf{z}}$, with $p_e$ the electric dipole moment per unit of length. The emitter is embedded in the host dielectric with relative permittivity $\varepsilon_{r,s}$ and is located at $(x,z) = (x_0, z_0)$ with $z_0 > 0$ [see Fig. S4]. For simplicity of modeling, the current density is independent of the $y$-coordinate so that the problem is two-dimensional.

Looking for a solution of the Maxwell's equations

$$\nabla \times \mathbf{E} = i\omega\mu_0\mathbf{H}, \qquad \nabla \times \mathbf{H} = -i\omega\varepsilon_0\varepsilon_{r,s}\mathbf{E} + \mathbf{j}_e, \qquad (S4)$$

of the form $\mathbf{H} = H_y(x,z)\hat{\mathbf{y}}$ it is found that for $z > 0$:

$$\nabla^2 H_y + \left(\frac{\omega}{c}\right)^2 \varepsilon_{r,s} H_y = -i\omega p_e \frac{\partial}{\partial x}\left[\delta(x-x_0)\delta(z-z_0)\right]. \qquad (S5)$$

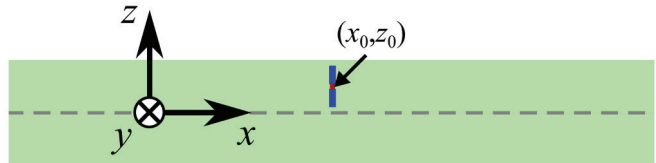

**Fig. S4.** A linearly polarized emitter is placed at the position $(x,z) = (x_0, z_0)$ above the graphene sheet.

In a bulk (unbounded) dielectric (without the graphene sheet) the emitted field is:

$$H_y^{inc} = i\omega p_e \frac{\partial}{\partial x}\left[-\frac{1}{4i}H_0^{(1)}\left(\frac{\omega}{c}\sqrt{\varepsilon_{r,s}}\,r\right)\right]$$
$$= -\omega p_e \frac{1}{2\pi}\int \frac{k_x}{2\gamma_s} e^{-\gamma_s|z-z_0|} e^{ik_x(x-x_0)} dk_x$$
, (S6)

where $H_0^{(1)}$ is the Hankel function of first kind and order zero, $r = \sqrt{(x-x_0)^2 + (z-z_0)^2}$, $\gamma_s = -i\sqrt{(\omega/c)^2 \varepsilon_{r,s} - k_x^2}$, and the integration range is over the entire real axis. The integral representation of the emitted field can be easily modified to take into account for the presence of the graphene sheet at $z=0$:

$$H_y = -\omega p_e \frac{1}{2\pi}\begin{cases} \int \frac{k_x}{2\gamma_s}\left(e^{-\gamma_s|z-z_0|} + \rho e^{-\gamma_s(z+z_0)}\right) e^{ik_x(x-x_0)} dk_x, & z > 0 \\ \int \frac{k_x}{2\gamma_s}\tau e^{\gamma_s(z-z_0)} e^{ik_x(x-x_0)} dk_x, & z < 0 \end{cases}$$
, (S7)

where $\rho = \rho(\omega, k_x)$ and $\tau = \tau(\omega, k_x)$ represent the (magnetic field) reflection and transmission coefficients for transverse magnetic (TM)-polarized waves, which are given in Sect. B. The x- and z- components of the electric field can be obtained from the Maxwell-Ampère equation so that:

$$E_x = \frac{ip_e}{\varepsilon_0 \varepsilon_{r,s}}\frac{1}{2\pi}\begin{cases} \int \frac{k_x}{2}\left(-\text{sgn}(z-z_0)e^{-\gamma_s|z-z_0|} - \rho e^{-\gamma_s(z+z_0)}\right) e^{ik_x(x-x_0)} dk_x, & z > 0 \\ \int \frac{k_x}{2}\tau e^{\gamma_s(z-z_0)} e^{ik_x(x-x_0)} dk_x, & z < 0 \end{cases}$$
, (S8)

$$E_z = \frac{p_e}{\varepsilon_0 \varepsilon_{r,s}}\frac{1}{2\pi}\begin{cases} \int \frac{k_x^2}{2\gamma_s}\left(e^{-\gamma_s|z-z_0|} + \rho e^{-\gamma_s(z+z_0)}\right) e^{ik_x(x-x_0)} dk_x, & z > 0 \\ \int \frac{k_x^2}{2\gamma_s}\tau e^{\gamma_s(z-z_0)} e^{ik_x(x-x_0)} dk_x, & z < 0 \end{cases}$$
. (S9)

## *D. Effect of the scattering objects*

We consider now that in addition to the graphene sheet a finite number of scatterers are positioned at $\mathbf{r} = \mathbf{r}_i$ ($i = 1, 2, ..., N$) (Fig. S5). The obstacles are assumed to be plasmonic thin strips (polarizable only along the vertical direction). The electric polarizability of the objects is roughly estimated (from an analogy with the polarizability of objects with a circular cross-section) to be $\alpha_{e,i} \approx (\pi w_i^2/2)(\varepsilon_{r,i} - \varepsilon_{r,s})/(\varepsilon_{r,i} + \varepsilon_{r,s})$, where $\varepsilon_{r,i}$ and $w_i$ are the relative permittivity and width (along z) of the i-th obstacle, respectively.

The effect of the strips on the electromagnetic field distribution is characterized using the formalism described in (*11*), which treats the obstacles as secondary sources of radiation. From the superposition principle, the total field is the superposition of the individual fields created by the emitter and obstacles in presence of the graphene sheet.

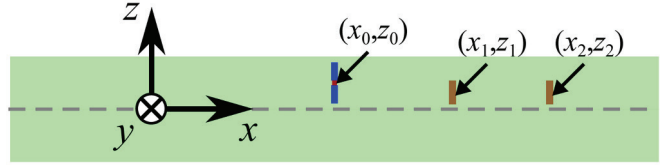

**Fig. S5.** Similar to Fig. S4, but with a few scattering objects placed at the positions $(x,z) = (x_i, z_i)$.

The local field in the vicinity of a given obstacle is the total electric field excluding the self-field produced by the obstacle. Thus, the local field on the i-th obstacle can be written as follows:

$$E_{\text{loc},i} = E_z^{\text{emitter}}(\mathbf{r}_i) + E_{z,i}^{\text{bs}}(\mathbf{r}_i) + \sum_{j \neq i} E_{z,j}(\mathbf{r}_i), \tag{S10}$$

where $E_z^{\text{emitter}}$ is the field radiated by the emitter in the presence of the graphene sheet (see Sect. C), $E_{z,i}^{\text{bs}}$ is the field backscattered by the graphene sheet due to the radiation of the i-th object, and $\sum_{j \neq i} E_{z,j}$ represents the field radiated by all the other obstacles.

Let $p_{e,i}$ be the unknown electric dipole moment (along z) induced on the i-th obstacle. Then, from Eq. (S9) the field backscattered by the graphene sheet can be written as $E_{z,i}^{\text{bs}}(\mathbf{r}_i) = C_s(z_i) p_{e,i}/(\varepsilon_0 \varepsilon_{r,s})$ with:

$$C_{\rm s}(z_i) = \frac{1}{2\pi} \int \frac{k_x^2}{2\gamma_{\rm s}} \rho e^{-\gamma_{\rm s} 2|z_i|} dk_x. \tag{S11}$$

Taking into account that for each scatterer the dipole moment is related to the local field as $p_{e,i} = E_{{\rm loc},i} \varepsilon_0 \varepsilon_{\rm r,s} \alpha_{e,i}$, it follows from Eq. (S10) that the unknown dipole moments satisfy:

$$\begin{pmatrix} C_{\rm S}(z_1) - \alpha_{e,1}^{-1} & C_{12} & \cdots \\ C_{21} & C_{\rm S}(z_2) - \alpha_{e,2}^{-1} & \cdots \\ \cdots & \cdots & \cdots \end{pmatrix} \begin{pmatrix} p_{e,1}/\varepsilon_0\varepsilon_{\rm r,s} \\ p_{e,2}/\varepsilon_0\varepsilon_{\rm r,s} \\ \cdots \end{pmatrix} = -\begin{pmatrix} E_z^{\rm emitter}(z_1) \\ E_z^{\rm emitter}(z_2) \\ \cdots \end{pmatrix}. \tag{S12}$$

The interaction constants $C_{i,j} = C_X(x_i - x_j; z_i, z_j)$ describe the coupling between the $i$-th and $j$-th obstacles and are given by [see Eq. (S9)]

$$C_X(x_i - x_j; z_i, z_j) = \frac{1}{2\pi} \int \frac{k_x^2}{2\gamma_{\rm s}} \left( e^{-\gamma_{\rm s}|z_i - z_j|} + \rho e^{-\gamma_{\rm s}(z_i + z_j)} \right) e^{ik_x(x_i - x_j)} dk_x. \tag{S13}$$

For simplicity, it is assumed in (S13) that all the obstacles are on the same side of the graphene sheet as the emitter, i.e. $z_i > 0$ ($i=1,\ldots,N$). Furthermore, the field radiated by the emitter in the presence of the graphene sheet can be written as $E_z^{\rm emitter}(x_i, z_i) = \frac{p_e}{\varepsilon_0\varepsilon_{\rm r,s}} C_X(x_i - x_0; z_i, z_0)$.

By solving Eq. (S12) one finds the dipole moments $p_{e,i}$ of the obstacles. The electromagnetic field produced by each of the obstacles can be calculated using Eqs. (S7-S9) with $p_e$ replaced by $p_{e,i}$. Finally, the total field is the superposition of the fields radiated by the emitter and by the secondary sources (i.e., the obstacles).

## E. Variation of the obstacle attenuation with the frequency and drift velocity

Figure S6 depicts the obstacle attenuation defined as $\left|E_x^{\text{no-obst}} / E_x^{\text{obst}}\right|$, with $E_x^{\text{obst}}$ the longitudinal electric field calculated beyond the obstacle position and $E_x^{\text{no-obst}}$ the field calculated at the same position but with a free-propagation path. As shown by the black curve in Fig. S6A the obstacle attenuation without a drift current is particularly strong near 16 THz.

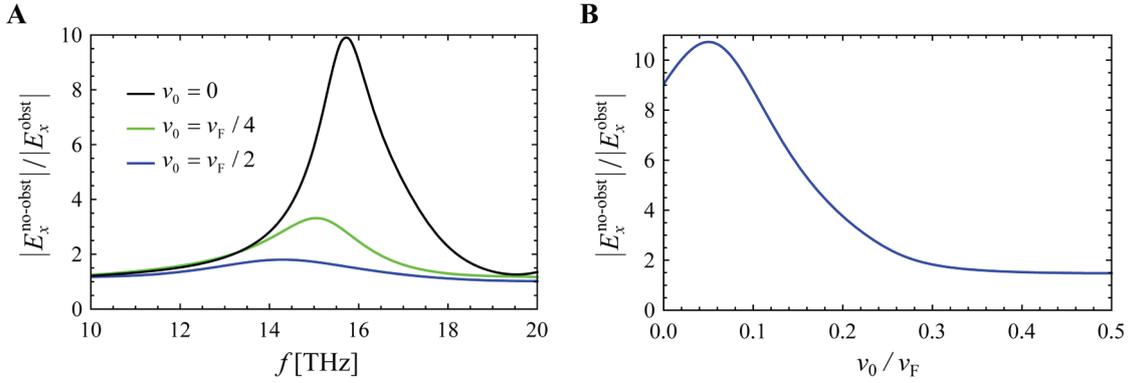

**Fig. S6.** Obstacle attenuation as a function of (**A**) the frequency for different drift velocities $v_0$, and (**B**) of the normalized drift velocity $v_0/v_F$ at the fixed frequency $f = 16$ THz. The emitter is placed at $(x_0, z_0) = (0, 10 \text{ nm})$ and the obstacle at $(x_1, z_1) = (150 \text{ nm}, 15 \text{ nm})$. The field is measured at the point $(x_p, z_p) = (300 \text{ nm}, 0)$. The obstacle has the same parameters as in the main text.

In agreement with the results of the main text, the attenuation due to the backscattering and absorption by the obstacle can be strongly reduced when the drift-current bias is applied to the graphene sheet. For example, for $v_0 = v_F/2$ (blue solid curve in Fig. S6A), the obstacle attenuation never exceeds 1.8 whereas without the drift current (black solid curve in Fig. S6A) it can exceed 10. Furthermore, as shown in Fig. S6B the obstacle attenuation at 16 THz is strongly suppressed for drift velocities larger than $0.2 v_F$.

# REFERENCES

1. Morgado, T. A.; Silveirinha, M. G. Negative Landau damping in bilayer graphene. *Phys. Rev. Lett.* **2017**, *119*, 133901.

2. Gonçalves, P. A. D.; Peres, N. M. R. *An introduction to graphene plasmonics*; World Scientific, Hackensack, NJ, 2016.

3. Wunsch, B.; Stauber, T.; Sols, F.; Guinea, F. Dynamical polarization of graphene at finite doping. *New J. Phys.* **2006**, *8*, 318.

4. Jablan, M.; Buljan, H.; Soljačić, M. Plasmonics in graphene at infrared frequencies. *Phys. Rev. B* **2009**, *80*, 245435.

5. Koppens, F. H. L.; Chang, D. E.; García de Abajo, F. J. Graphene Plasmonics: A Platform for Strong Light-Matter Interactions. *Nano Lett.* **2011**, *11*(8), 3370-3377.

6. Lin, X.; et al. All-angle negative refraction of highly squeezed plasmon and phonon polaritons in graphene-boron nitride heterostructures. *Proc. Natl. Acad. Sci. USA* **2017**, *114*, 6717–6721.

7. Van Duppen, B.; Tomadin, A.; Grigorenko, A. N.; Polini, M. Current-induced birefrigent absorption and non-reciprocal plasmons in graphene. *2D Mater.* **2016**, *3*, 015011.

8. Wenger, T.; Viola, G.; Kinaret, J.; Fogelström, M.; Tassin, P. Current-controlled light scattering and asymmetric plasmon propagation in graphene. *Phys. Rev. B* **2018**, *97*, 085419.

9. Sabbaghi, M.; Lee, H.-W.; Stauber, T.; Kim, K. S. Drift-induced modifications to the dynamical polarization of graphene. *Phys. Rev. B*, **2015**, *92*, 195429.

10. Hanson, G. W. Dyadic Green's functions and guided surface waves for a surface conductivity model of graphene. *J. Appl. Phys.* **2008**, *103*, 064302.

11. Silveirinha, M. G.; Medeiros, C. R.; Fernandes, C. A.; Costa, J. R. Resolving subwavelength objects with a crossed wire mesh superlens operated in backscattering mode. *New J. Phys.* **2011**, *13*, 053004.
10